\documentclass[useAMS,usenatbib]{mn2e}
\usepackage{times}
\usepackage{amsmath}
\usepackage{amssymb,amsfonts}
\usepackage{graphicx}
\usepackage{verbatim}
\usepackage{psfig}
\usepackage{epsf}

\def\ltsima{$\; \buildrel < \over \sim \;$}
\def\simlt{\lower.5ex\hbox{\ltsima}}
\def\gtsima{$\; \buildrel > \over \sim \;$}
\def\simgt{\lower.5ex\hbox{\gtsima}}
\def\kpc{{\rm\,kpc}}

\def\deg{^\circ}

\def\s{\ifmmode \widetilde \else \~\fi}
\def\={\overline}

\def\spose#1{\hbox to 0pt{#1\hss}}

\def\ie{{ i.e.,\ }}
\def\lta{\mathrel{\spose{\lower 3pt\hbox{$\mathchar"218$}}
     \raise 2.0pt\hbox{$\mathchar"13C$}}}
\def\gta{\mathrel{\spose{\lower 3pt\hbox{$\mathchar"218$}}
     \raise 2.0pt\hbox{$\mathchar"13E$}}}
\def\Dt{\spose{\raise 1.5ex\hbox{\hskip3pt$\mathchar"201$}}}    
\def\dt{\spose{\raise 1.0ex\hbox{\hskip2pt$\mathchar"201$}}}    

\def\dotsfill{\leaders\hbox to 1em{\hss.\hss}\hfill}

\title
[The AAT/WFI survey of the Monoceros Ring and Canis Major Dwarf galaxy: II. from {\it l} = (280 - 025)$\deg$]
{The AAT/WFI survey of the Monoceros Ring and Canis Major Dwarf galaxy: II. from {\it l} = (280 - 025)$\deg$}
\author[Blair Conn et al.]
       {Blair C. Conn$^{1,2}$, Richard R. Lane$^1$, Geraint F. Lewis$^1$, Mike J. Irwin$^3$, 
       \newauthor Rodrigo A. Ibata$^4$, Nicolas F. Martin$^4$, Michele Bellazzini$^5$,
       \newauthor \& Artem V. Tuntsov$^{1,7}$\\
$^{1}$Institute of Astronomy, School of Physics, A29, University of Sydney, NSW 2006, Australia:\\
Email \tt{bconn@eso.org}\\
$^{2}$European Southern Observatory, Alonso de Cordova 3107, Vitacura, Santiago, Chile\\
$^{3}$Institute of Astronomy, Madingley Road, Cambridge, CB3 0HA, U.K.\\
$^{4}$Observatoire de Strasbourg, 11, rue de l'Universit\'e, F-67000, Strasbourg, France\\
$^{5}$INAF - Osservatorio Astronomico di Bologna, Via Ranzani 1, 40127, Bologna, Italy\\
$^{6}$Anglo-Australian Observatory, Epping, NSW, 1710, Australia\\
$^{7}$Sterberg Astronomical Institute, Moscow, Russia\\
$^{8}$Institute for Astronomy, University of Edinburgh, Royal Observatory, Blackford Hill, Edinburgh, EH9 3HJ, U.K.}

\begin{document}

\date{\today \hspace{10pt}(Version 3.0)}

\pagerange{\pageref{firstpage}--\pageref{lastpage}} \pubyear{2008}

\def\LaTeX{L\kern-.36em\raise.3ex\hbox{a}\kern-.15em
    T\kern-.1667em\lower.7ex\hbox{E}\kern-.125emX}

\newtheorem{theorem}{Theorem}[section]

\label{firstpage}

\maketitle

\begin{abstract}
This paper concludes a systematic search for evidence of the Monoceros
Ring and Canis Major dwarf galaxy around the Galactic Plane. Presented
here are the results for the Galactic longitude range of {\it l} =
(280 - 025)$\deg$. Testing the claim that the Monoceros Ring encircles
the entire Galaxy, this survey attempts to document the position of
the Monoceros Ring with increasing Galactic longitude. Additionally,
with the discovery of the purported Canis Major dwarf galaxy,
searching for more evidence of its interaction is imperative to
tracing its path through the Galaxy and understanding its role in the
evolution of the Milky Way. Two new detections of the Monoceros Ring
have been found at ({\it l, b}) = (280,$+$15)$\deg$ and
(300,$+$10)$\deg$. Interestingly, in general there seem to be more
detections above the Plane than below it; in this survey
around $\frac{2}{3}$ of the firm Monoceros Ring detections are in the
North. This coincides with the Northern detections appearing to be
qualitatively denser and broader than their Southern counterparts. The
maximum of the Galactic Warp in the South is also probed in this
survey. It is found that these fields do not resemble those in the
Canis Major region suggesting that the Warp does not change
the shape of the CMD as is witnessed around Canis Major. The origins
and morphology of the Monoceros Ring is still elusive primarily due to its
enormous extent on the sky. Continued probing of the Galactic Outer
Disc is needed before a consensus can be reached on its nature.

\end{abstract}

\begin{keywords}
Galaxy:\hspace{2pt}formation -- Galaxy:\hspace{2pt}structure -- galaxies:\hspace{2pt}interactions
\end{keywords}

\begin{table*}
\centering
\caption{Summary of the observations  of Monoceros Ring/Canis Major Tidal Stream with
  the AAT/WFI, ordered in ascending Galactic longitude ({\it l}). The
  number of CCDs available during the different runs varies and thus
  has effected the total area or Field of View observed.}
\begin{tabular}{lcccccccl} \hline \hline
Fields ({\it l,b})$\deg$ & Regions per field & Average Seeing (arcsec)
 & Total Area ($deg^2$)  & Monoceros Ring  & Average E(B-V) &
 Date (dd/mm/yy)\\ \hline
 (280,$-$15)$\deg$    & 4  & 1.3 & 1.21    & No  & 0.128 & 01/02/04\\
 (280,$+$15)$\deg$    & 3  & 1.0 & 0.93    & Yes    & 0.083 & 25,30/01/04\\
 (300,$-$20)$\deg$    & 1  & 1.3 & 0.3     & No  & 0.109 & 31/01/04\\
 (300,$+$10)$\deg$    & 3  & 1.2 & 0.93    & Yes    & 0.171 & 25/01/04\\
 (340,$+$20)$\deg$    & 4  & 2.8,1.6 &0.91 & No     & 0.095 & 15-16/08/05\\
 (350,$-$20)$\deg$    & 4  & 2.2 & 0.91    & No     & 0.055 & 15/08/05\\
 (350,$+$20)$\deg$    & 4  & 1.4 & 0.91    & No     & 0.112 & 16/08/05\\
 (025,$-$20)$\deg$    & 5  & 2.0 & 1.14    & No     & 0.137 & 15-16/08/05\\
 (025,$+$20)$\deg$    & 5  & 2.0 & 1.14    & No     & 0.098 & 15-16/08/05\\
\hline\hline
\end{tabular}
\label{ObsTable}
\end{table*}

\section{Introduction}
The Monoceros Ring (MRi), discovered in 2002
\citep{2002ApJ...569..245N} has now been traced around the Galaxy from
{\it l} = 75 - 260$\deg$, as shown through the Sloan Digital Sky
Survey \citep{2002ApJ...569..245N}, Two-micron All Sky Survey
\citep{2003ApJ...594L.115R}, Isaac Newton Telescope Wide Field Camera
Survey \citep{2003MNRAS.340L..21I} and the Anglo-Australian Telescope
Wide Field Imager Survey \citep{2005MNRAS.362..475C}. Continuing
around the Galactic plane, this survey extends these previous results
to complete the first Wide Field Imager survey of the MRi around the
Galaxy that began with the INT/WFC in
\citet{2005MNRAS.362..475C}. Studies into this structure have been
discussed in Paper I of this series, \citep[][]{2007MNRAS.376..939C},
and references therein. Additional to this, an RR-Lyrae survey of the
Galactic Halo using QUEST data has also revealed the presence of the
MRi and investigated the overdensity in Canis Major
\citep[][]{2006AJ....132..714V,2007IAUS..241..359M}.

\begin{figure}
\centerline{
\psfig{figure=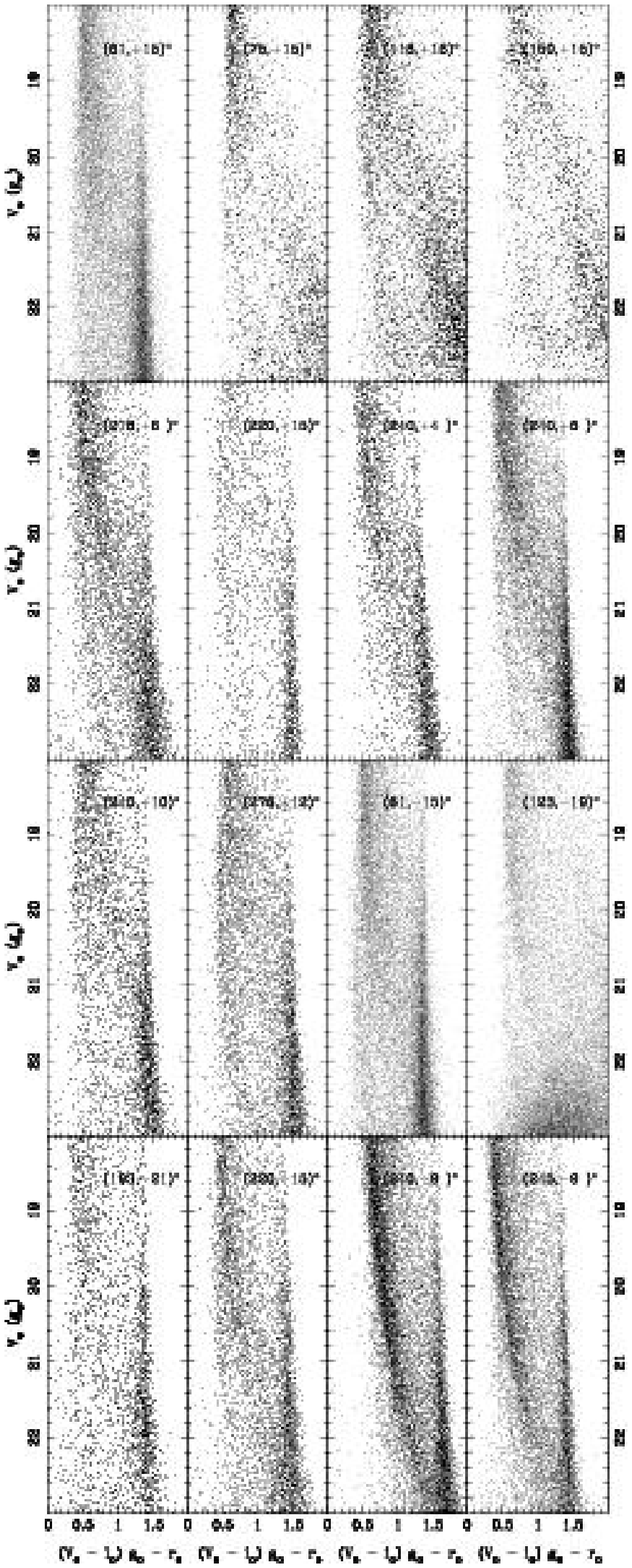}}
\caption[]{Visual summary of all the previous Monoceros Ring detections from
  the INT/WFC survey \citep{2005MNRAS.362..475C} and AAT/WFI survey (Paper I) of the outer Disc. The
  Colour-Magnitude diagrams are of two types {\it V,i} and {\it
  g,r}. For more detailed analysis of these fields and the reported
  detections see the relevant articles.
\label{figmonster}}
\end{figure}

Residing in the Thick Disc of the Milky Way (MW), the MRi is revealed
only by obtaining deep photometry of large patches of sky, typically
greater than 1 square degree. In this preliminary first pass of the
MW, the Thick Disc was sampled at Galactic latitudes nominally between
{\it b} = $\pm$10$\deg$-20$\deg$ and about every 20 degrees in
Galactic longitude. To date, the entire survey has strong detections
of the MRi in 14 regions with 3 additional tentative detections out of
25 regions observed. It has been found on both sides of the Galactic
plane at Galactic latitudes from 4$\deg$ - 20$\deg$ and its extent
away from the plane is as yet undetermined although SDSS results
suggest that it could be as high as +30$\deg$
\citep{2007ApJ...658..337B}. Numerical simulations of the MRi as a
tidal stream predict it to have multiple wraps around MW, although the
current dataset cannot distinguish between different aspects of the
stream nor whether the different detections are part of a coherent
structure. Figure~\ref{figmonster} shows the previous detections of
the MRi as reported in \citet{2005MNRAS.362..475C} and
\citet{2007MNRAS.376..939C}. Figure 2 of Paper I shows how these CMDs
can be interpreted by showing the approximate location of the Thin,
Thick and Halo stars in the field. Figure 5 of Paper I illustrates
which components are being referenced when discussed in the text. Each
of the fields in this figure are pixelated Colour-Magnitude
diagrams. The pixel values represent the square-root of the number of
stars in that pixel. Ordered by increasing Galactic Longitude,
Figure~\ref{figmonster} tentatively shows the changing strength and thickness of
the MRi around the Plane. Qualitatively, the strength of the MRi can
be seen in comparison with the MW components. Additionally, the only
apparent difference between Northern and Southern detections is
perhaps that the Southern MRi features appear qualitatively narrower than their
Northern counterparts. There is no clear explanation as to why
this is the case.

While there is more and more evidence regarding the true extent of
this structure, there is very little information concerning many of
its generic properties. As such, no direct measurement of the density
profile of any part of the stream around the sky has been made, nor a
complete understanding of its true extent on the sky. In the region
covered by the SDSS, \citet{2008ApJ...673..864J} and
\citet{2008arXiv0804.3850I}, report on the presence of the MRi with
regard to its number density, metallicity and kinematics. A clear
overdensity of stars can be seen at a distance of 16 kpc
\citep{2008ApJ...673..864J} and \citet{2008arXiv0804.3850I} reports a
mean metallicity of [Fe/H] = -0.95 with a scatter of around 0.15
dex. Kinematically, they show a spread of velocities rotating
consistently faster than the Local Standard of Rest and in accordance
with the predictions of ~\citet{2005ApJ...626..128P}.

The only possible candidate for the stream's progenitor is an
overdensity of stars found in Canis Major but possible confusion with
the Galactic Warp has created doubts on this detection. Numerical
simulations created using the properties of these stars have predicted
the location and extent of the MRi with good accuracy and so adds
support to the dwarf galaxy scenario. This on-going debate centres on
whether observations of the Canis Major overdensity conform to known
Galactic structure, such as the Warp, or can be considered truly
additional. Paper I of this series outlines some of the possible
inconsistencies between predicted properties of the Galactic Warp and
direct observations of these structures. In response to this
\citet{2007arXiv0707.4440L} have presented an explanation relying on
only first order Galactic structure. Further refinement of the
properties of the MRi are needed to determine whether CMa is the most
likely candidate as its progenitor.

\section{Observations and Reduction}\label{obs}
The Anglo-Australian Telescope Wide Field Imager (AAT/WFI)
at Siding Spring Observatory in New South Wales, Australia was used to
conduct the current survey. The AAT/WFI is mounted at prime focus with a field
of view of approximately 33 arcminutes on a side. It consists of eight
4k$\times$2k  CCDs with a pixel scale of 0.2295 arcsec per pixel. 

The observations were taken over three observing runs, the first on
the 22-25 January 2004, the second on 30, 31st of January and 01
February 2004 with the third on the 14, 15 and 16th of August 2004. To
minimise the fringing effects that are present when observing with
other filters we employed the {\it g} (WFI SDSS \#90) 
and {\it r} (WFI SDSS \#91) filters. Exposure times used were a single 600
second exposure in {\it g} and two 450 second exposures in {\it r}. Twilight flats along with
bias and dark frames were used for calibration, and Landolt Standard Star
fields were observed roughly every two hours. Data reduction was
performed using the Cambridge Astronomical Survey Unit (CASU) Pipeline
~\citep{2001NewAR..45..105I}, a thorough description of this process
and the necessary calibrations are outlined in Paper I of this series.

This paper presents the final section of the survey using the AAT/WFI
which observed fields from {\it l} = (280 - 25)$\deg$
across the Galactic bulge. This is in addition to Paper I which
covered fields in the regions {\it l} = (195 - 276)$\deg$. Nine fields have been observed, and in
general each field is approximately one square degree or four WFI
pointings. A summary of the results of this survey is shown in
Table~\ref{ObsTable}.

\section{Analysis}\label{analysis}

The magnitude completeness of the data has been estimated in the same manner as described
in Paper I. In essence, this involves using overlapping regions of the
observed fields to determine the completeness. The field at ({\it l,b}) = (300,-20)$\deg$ has only
one pointing and so with this approach no completeness estimate is
possible. Table~\ref{CompTable} presents the completeness profiles
of each field based on the equation 
\begin{equation}\label{eqncomplete}
  CF = \frac{1}{1 + e^{(m - m_c)/ \lambda}}\\
\end{equation}
  
\begin{table}
\centering
\caption{Parameters used to model the completeness of each field,
  ordered in ascending Galactic longitude ({\it l}). $m_c$ is the
  estimated 50\% completeness level for each filter with $\lambda$ describing the
  width of the rollover function (see Equation~\ref{eqncomplete} and
  further details in Paper I). }
\begin{tabular}{||l|c|c|c||} \hline \hline
Fields \it{(l,b)$\deg$} & m$_c$  ($g_\circ$)  & m$_c$ ($r_\circ$)  & $\lambda$ \\ \hline
(280, $-$15)$\deg$ & 22.40 & 21.40 & 0.55\\
(280, $+$15)$\deg$ & 23.85 & 22.60 & 0.60\\
(300, $-$20)$\deg$ &\multicolumn{3}{c}{no estimate possible}\\
(300, $+$10)$\deg$ & 23.30 & 22.40 & 0.30\\
(340, $+$20)$\deg$ & 21.80 & 20.70 & 0.45\\
(350, $-$20)$\deg$ & 23.10 & 22.30 & 0.30\\
(350, $+$20)$\deg$ & 23.40 & 22.40 & 0.75\\
(025, $-$20)$\deg$ & 22.60 & 21.70 & 0.60\\
(025, $+$20)$\deg$ & 22.60 & 21.70 & 0.60\\
\hline	
\end{tabular}
\label{CompTable}
\end{table}

Estimating the completeness provides a way to evaluate the quality of
the data and helps to determine the reliability of the detections in
the faint end of the Colour-Magnitude diagram (CMD). While in Paper I the completeness profile
was used when making signal-to-noise estimates of the stream, the data here are not
of sufficient quality to allow such a measurement. This is because
many of these fields only have two or three pointings per region (fewer stars)
coupled with poor seeing leading to a relatively bright limiting
magnitude. With these factors, the MRi is not as clearly above the
noise as in Paper I.

Following the method employed by \citet{2003MNRAS.340L..21I},
\citet{2005MNRAS.362..475C} and Paper I,  we have used a main sequence
fiducial to estimate the distance to the features seen in the
CMDs. For a complete explanation of the process
and errors involved see $\S$4.1.2 of Paper I. The furthest distance to
which this method can find the MRi is difficult to
estimate. The number density of MRi stars per field and the distance to
the MRi are obviously important to whether a detection is
made. Additionally, the quality of the data in those fields will again
directly influence the likelihood of a detection. Poor seeing and
insufficient sky coverage could easily effect the ability of this
method to make a successful detection. The findings of this survey
suggest that if the MRi is within $\sim$20 kpc it will be
detectable. Beyond this, it is highly dependent on the strength of MRi
in the CMD and only one field has the MRi placed greater than 20
kpc. A possible reason for this is that a detection at a distance of
20 - 30 kpc involves a shift of 1.5 - 2.2 magnitudes from the base
position at 11 kpc ({\it g} = 19.5). The turn-off of the shifted main
sequence is now located at $\sim$22 magnitude where the photometric
errors are starting to increase and thereby spread out the main
sequence. In the absence of very deep or wide surveys this apparent
limit of $\sim$20 kpc may remain the practical limit for finding the MRi.

\subsection{Comparisons with the Besan\c{c}on Model}\label{comparisons}
The basic methodology we have employed when searching for additional structures in
the outer Disc of the Galaxy is to make direct comparisons with the
Besan\c{c}on model which purports to predict its properties. While
this approach is not favoured by some it has a few
advantages. Firstly, and quite importantly it allowed the survey to be
completed in a reasonable time frame. Adding an extra filter, such as
a U band or i band filter, would have dramatically increased the time
needed. Secondly, the dynamic range of the survey means that the
brighter end of the survey can test the predicted properties of the
bulk Milky Way components while the fainter end tests the outer Disc
region. Since the MRi is only distinguishable in the Thick Disc/Halo
part of the CMD, searching for its presence relies on looking at the
fainter end of the CMD. The Canis Major dwarf galaxy feature as
discussed in Paper I, is located more or less in the Thin Disc
component. Indeed, since most of the debate concerning CMa revolves around
whether the CMDs observed in the CMa region are explainable in terms
of purely Disc components or whether an extra component exists in the
same Colour-Magnitude space. It is for the latter possibility that the
distance to the edge of the Thin Disc component has been determined for the
fields in this part of the survey. Checking their position with respect
to the model provides an opportunity to assess whether it is different
and perhaps could be related to the CMa overdensity. For some fields, measuring the faint
edge of the Thin disc cut-off has not been possible due to the CMDs
not showing a clear edge. For these fields, the distance to the bright edge of the Thin
Disc region has been found. So for each field there are three
possible structures to be examined: the faint MRi component, which may
represent additional MW substructure; the faint edge
of the Thin Disc, which may represent a mis-identified CMa-type population as per Paper I;
or the upper edge of the Thin Disc, which tests the model in these
directions. The results of these parameters are presented in Table~\ref{ResultsTable}.

\subsection{Survey Fields}\label{Survey Fields}

\begin{figure}
\centerline{
\psfig{figure=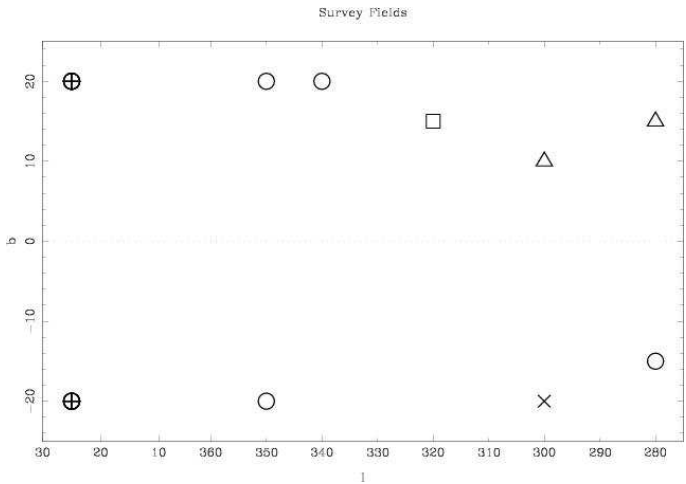,angle=270,width=9cm}}
\caption[]{This figure shows the location of the survey fields
  presented in this paper. The symbols denote the number
  of pointings per field. The circles with plus signs represent five
  pointings; the empty circles, four pointings; the triangles, three
  pointings; the square, two pointings; and the cross one
  pointing. Each field was intended to have four-five pointings,
  however, weather and time constraints have resulted in many fields containing
  less. The survey was originally designed to contain a more complete
  coverage of the Galactic plane, however, weather again significantly
  limited the number of photometric or near-photometric fields with
  useful data. The resultant fields have been selected based on data
  quality and information content. While ideally four pointings would
  correspond to about 1 square degree of sky, observations undertaken
  during August 2004 had only 6 of 8 CCDs available, and those taken
  during February 2006 had only 7 of 8 CCDs available. This has
  limited the amount of sky surveyed, despite attempts to reduce its
  impact. The actual Field of View observed can be found in
  Table~\ref{ObsTable} under the Total Area column.
\label{figsurvey}}
\end{figure}

\begin{table}
\caption{\small Summary of the observations  of Monoceros Ring/Canis Major Tidal Stream with
  the AAT/WFI, ordered in ascending Galactic longitude ({\it l}). The offset is measured in magnitudes from the zero offset position of the \citet{2002ApJ...569..245N} detections at 11 kpc.}
\begin{tabular}{lcccccc} \hline \hline
Fields ({\it l,b})$\deg$ & MRi offset  & MRi dist       & MW/CMa offset & MW/CMa dist \\
                      &  (mag) & (kpc) & (mag)  & (kpc) \\ \hline
 (280,$-$15)$\deg$& - & -   & -      & -   \\
 (280,$+$15)$\deg$& 0.0 & 11.0   & -      & -   \\
 (300,$-$20)$\deg$& - & -   & $-$0.8 & 7.6 \\
 (300,$+$10)$\deg$& 0.8 & 15.9   & $-$0.8 & 7.6 \\	
 (340,$+$20)$\deg$& -   & -      & $-$0.2 & 10.0\\
 (350,$-$20)$\deg$& -   & -      & $-$0.5 & 8.7 \\
 (350,$+$20)$\deg$& -   & -      & $+$0.2 & 12.1\\
 (025,$-$20)$\deg$& -   & -      & $-$2.0 & 4.4 \\
 (025,$+$20)$\deg$& -   & -      & $-$2.2 & 4.0 \\
\hline\hline
\end{tabular}
\label{ResultsTable}
\end{table}

The location of each field, in Galactic coordinates, is shown graphically in Figure~\ref{figsurvey}.
Each field is presented in the following sections showing the CMDs
with the appropriate main sequence type overlay as taken from the
original \citet{2002ApJ...569..245N} detection and described in Paper
I. All magnitude offsets of the main sequence overlay are with respect
to this \citet{2002ApJ...569..245N} detection at 11.0
kpc. Table~\ref{ResultsTable} summarizes the outcome of this study and
uses the same formatting as in Paper I. It should be noted though that
this paper does not find evidence of the Canis Major dwarf and the
final column of Table~\ref{ResultsTable} simply presents where the
fiducial main sequence has been placed. In general, for this part of
the survey, the Besan\c{c}on model is well matched to the
data and as such the dominant main sequence is easily attributed to
known Galactic structure. The
CMDs that we have used are density maps of the
underlying distribution. Each pixel is the square root of the number
of stars in that part of the CMD. This method provides better contrast of the structures
especially in regions if high stellar density. A presentation of all the
fields from previous AAT and INT surveys in which the Monoceros Ring is present can be seen in
Figure~\ref{figmonster}. In the following sections we will provide the
distance estimates to the major features present in each CMD from this
part of the survey with an analysis of these results being presented
in the Discussion ($\S$\ref{discussion}).

\subsubsection{Fields \bf$(280,-15)^\circ$}\label{280m15des}
\begin{figure}\centering
\includegraphics[width=82mm,angle=270]{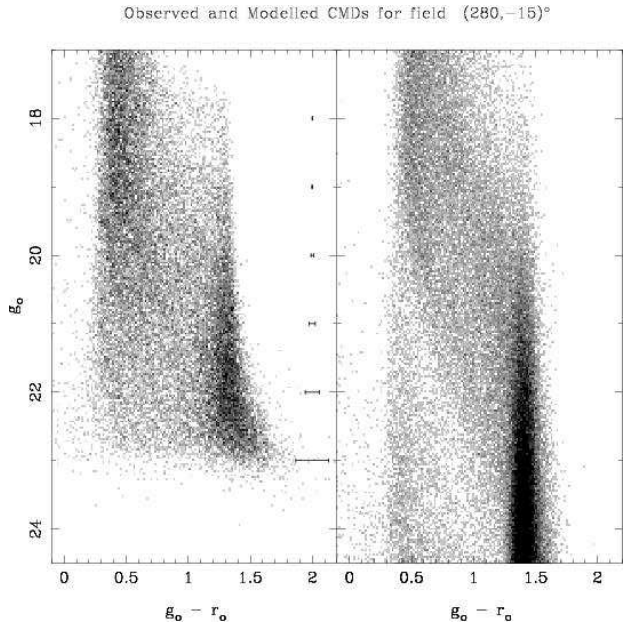}
\caption[]{Hess plot  of ({\it l},{\it b}) = (280,-15)$\deg$ and the
  corresponding Besan\c{c}on model.  A Hess plot is created by
  pixelating the Colour-Magnitude diagram and generating a grayscale
  on the basis of the square root of the pixel number density.  The
  same process is applied to both the data (Left panel) and
  model (Right panel).  The synthetic galaxy model was generated via
  the Besan\c{c}on online galaxy model
  website (http://bison.obs-besancon.fr/modele/). The distance
  interval applied to the model is a line-of-sight from the Sun out to
  100 kpc. This ensures that no artificial cuts can enter into the
  CMDs via distance effects. The model is selected
  in $g,r$ in the CFHTLS system and the converted to $g,r$ of the
  AAT/WFI via the colour conversions discussed in Section 4.1 of Paper
  I. 
\label{fig280m15}}
\end{figure}

The (280,-15)$\deg$ field (Figure~\ref{fig280m15}) is approximately 40 degrees from the
purported dwarf galaxy in Canis Major and the features here seem less
defined than in nearer fields. This is perhaps due to slight
differences in the photometric solution for each frame combined with
the brighter limiting magnitude. The strong main sequence seen in
({\it l,b}) = (273,-9)$\deg$ (Figure 20 of Paper I) is
not seen here although the increase in latitude away from the Galactic
plane could account for this change. Deeper imaging of this region is necessary to
confirm the lack of the CMa feature and to investigate the slight excess of stars in
the region {\it g$_{\circ}$} $>$ 21 and ({\it g - r})$_{\circ}$ $<$ 1.0.

\subsubsection{Fields \bf$(280,+15)^\circ$}\label{280p15des}
\begin{figure}\centering
\includegraphics[width=82mm,angle=270]{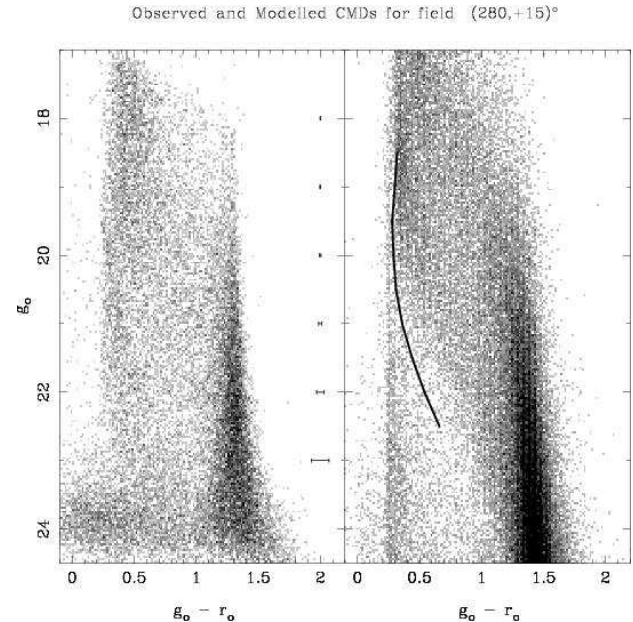}
\caption[]{Hess plot  of  ({\it l},{\it b}) = (280,+15)$\deg$ and the
  corresponding Besan\c{c}on model. The figure is otherwise the same
  as Figure~\ref{fig280m15}. The main sequence fitted here for the
  Monoceros Ring is offset by 0.8 magnitudes. The heliocentric
  distance related to this offset is then 15.9 \kpc. No error or
  signal to noise estimate has been derived for this feature.
\label{fig280p15}}
\end{figure}

The (280,+15)$\deg$ field (Figure~\ref{fig280p15}) is similar to its corresponding field
below the plane at (280,-15)$\deg$.  The comparison field from
the Besan\c{c}on model is presented here with the fiducial main
sequence at the location of the additional main sequence present in
the data. This main sequence has been interpreted as the Monoceros
Ring. Interestingly, the MRi feature in this field is more extended
than in others. The stream is perhaps extended or wrapped in this part
of the sky or the MW components here have different strengths than the
Besan\c{c}on model predicts. A mix of the two is also possible. The
fiducial shown marks the brighter edge of this feature. With only a small shift from the
nearby detection at (276,+12)$\deg$ (see Figure~\ref{figmonster}, the offset used for this feature is 0.8
magnitudes corresponding to 15.9 \kpc\ heliocentrically. In comparison
to the detection at (276,+12)$\deg$, this is about 4 \kpc\ further
away.  The lower edge of this feature is approximately 0.5 magnitudes
fainter and thus would be estimated at around 20 \kpc. No attempt has been made to estimate the width of this feature.
 
\subsubsection{Fields \bf$(300,-20)^\circ$}\label{300m20des}
\begin{figure}\centering
\includegraphics[width=82mm,angle=270]{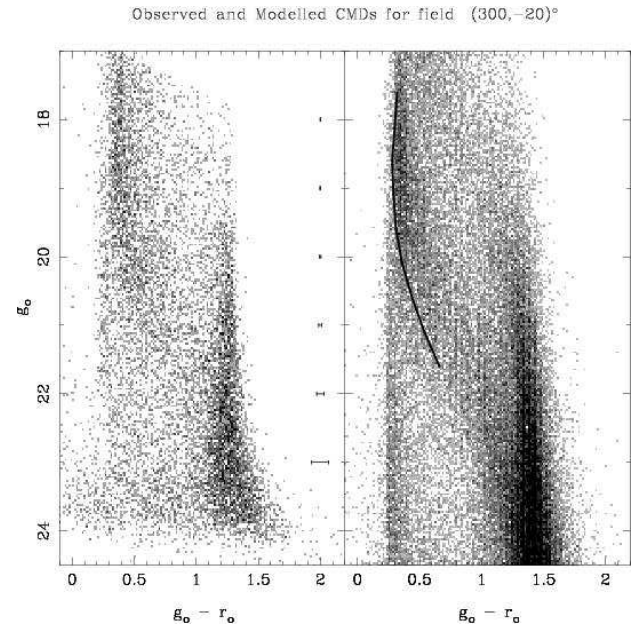}
\caption[]{Hess plot  of  ({\it l},{\it b}) = (300,-20)$\deg$ and the
  corresponding Besan\c{c}on model. The figure is in the same format
  as Figure~\ref{fig280m15}. The main sequences fitted here are offset
  by -0.8 magnitudes. The heliocentric distance of this offset is 7.6
  \kpc. The similarity with the synthetic CMD suggests this main
  sequence in the data is associated with Galactic disc.
\label{fig300m20}}
\end{figure}

At (300,-20)$\deg$ (Figure~\ref{fig300m20}), the main sequence crossing the middle of the CMD
is well matched by the synthetic CMD and corresponds to the location of
the Disc stars seen in the model. The overlay is offset at
-0.8 magnitudes or $\sim$7.6 \kpc. There is perhaps a main sequence
belonging to the MRi at the faint blue end of the CMD however the
model does indicate that some stars should be expected in that
location. Given the overall noisy quality of the CMD, no attempt is
made to identify whether those stars may belong to the MRi. The strong
main sequence defined by the fiducial is a good match with
the model and thus is most likely of Galactic origin.

\subsubsection{Fields \bf$(300,+10)^\circ$}\label{300p10des}
\begin{figure}\centering
\includegraphics[width=82mm,angle=270]{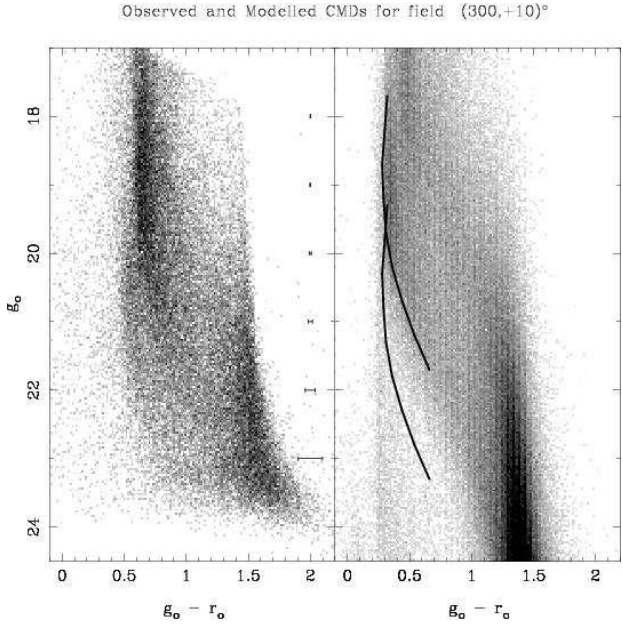}
\caption[]{Hess plot  of  ({\it l},{\it b}) = (300,+10)$\deg$ and the
  corresponding Besan\c{c}on model. The figure is otherwise the same
  as Figure~\ref{fig280m15}. The main sequences fitted here are offset
  by -0.8 and 0.8 magnitudes. The heliocentric distance of these
  offsets are 7.6 and 15.9 \kpc. The Galaxy component is related to
  the closer feature and the MRi to the more distant feature. Due to the data having been shifted
  by 0.1 magnitudes in {\it r} to align the CMD in colour, the
  distances have an additional source of uncertainty.
\label{fig300p10}}
\end{figure}

The (300,+10)$\deg$ field (Figure~\ref{fig300p10}) contains an obvious additional main
sequence more distant than the expected Milky Way component. The
original data for this field was slightly misaligned in colour after
all the photometric calibrations were applied. To try to
ensure the smallest shift possible when correcting this, the r magnitudes
have all been shifted by $+$0.1 magnitudes. Taking the field
without any differential extinction and shifting the others to match
aligns the final CMD in the correct colour range and allows the main sequence overlay to be used
to estimate the distance. Of course, shifting the data in this manner
weakens the accuracy to which we can determine the distance. While the
shift was small, all the distances reported for this field can only be
seen as indicative and do not have the accuracy as reported in the
other fields of the survey. The two overlays are offset by -0.8
magnitudes for the brighter main sequence and 0.8 for the fainter main
sequence. These result in distance estimates of $\sim$7.6\kpc\ and
$\sim$15.9 \kpc\ respectively. The stronger main sequence is clearly
related to the Galaxy given the good correlation with the model.

\subsubsection{Fields {\bf$(340,+20)^\circ$}}\label{340p20des}
\begin{figure}\centering
\includegraphics[width=82mm,angle=270]{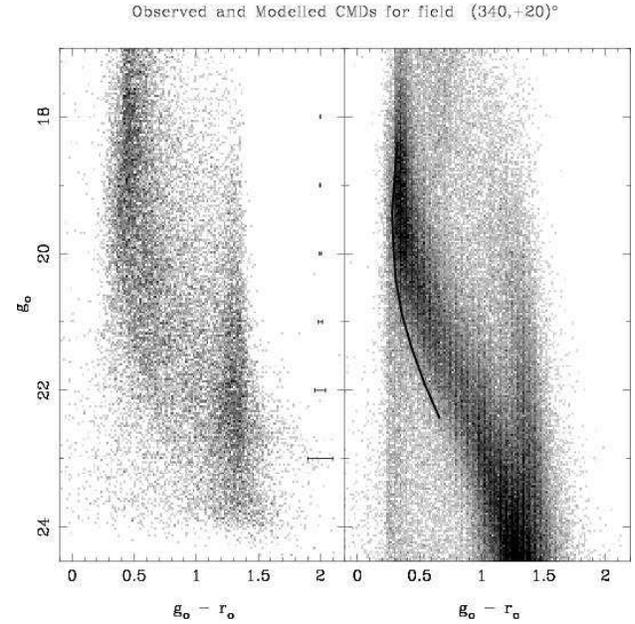}
\caption[]{ As for Figure~\ref{fig280m15}, Hess plot of ({\it l},{\it
  b}) = (340,+20)$\deg$. The offset is placed at -0.2 magnitudes, or
  10.0 \kpc\ heliocentrically and is clearly associated with the MW
  component in the model. The original CMD was slightly offset in
  colour and this has been corrected with a small shift of 0.1
  magnitudes in {\it r} towards the red. The distance estimates
  becomes less accurate due to this shift. 
\label{fig340p20}}
\end{figure}

The data in this field (Figure~\ref{fig340p20}) is a combination of two pointings which
resulted in different limiting magnitude when calibrated. This could
partly contribute to the lack of coherence in the data toward the
limiting magnitude of the shallower sample ({\it g$_\circ$}$\sim$22.5). When combining the two
datasets the selection criteria has been tightened; in the other
fields, if the object is classified a star in one filter and only possibly
a star in the other it is accepted. With this CMD, only if in both
filters the object is classified as a star has it been plotted. This
was done to try and remove some of the additional noise in the CMD. Additionally however, an
alignment in colour by $\sim$0.1 magnitudes redward was also required. This will impact on the accuracy of any distance estimates
of structures within this field. The overlay is fitted to the lower
extreme of the Milky Way main sequence and is a good match to the
predictions of the model. 

\subsubsection{Fields \bf$(350,-20)^\circ$}\label{350m20des}
\begin{figure}\centering 
\includegraphics[width=82mm,angle=270]{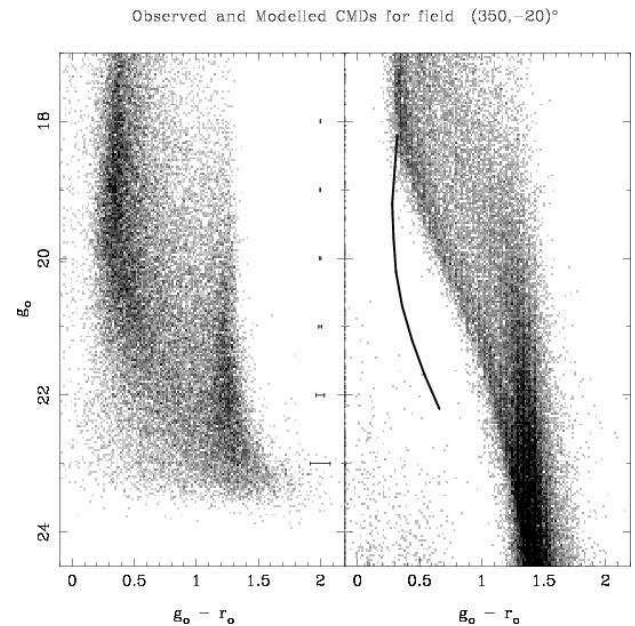}
\caption[]{Hess  plots of  ({\it l},{\it b}) = (350,-20)$\deg$. The
overlay is placed at -0.5 magnitudes or 8.7 \kpc\ heliocentric. The
model clearly has problems with this direction on the sky and any
differences are not expected to be real. The data does not seem to
have an MRi-like component.
\label{fig350m20}}
\end{figure}

This field (Figure~\ref{fig350m20}) shows a broad main sequence with an overlay placed with an
offset of -0.5 magnitudes (8.7 \kpc\ heliocentric distance). There is an
obvious problem with the predictions of the model. In the following
field, this problem was avoided by locating a field nearby which
reproduced an acceptable CMD. Unfortunately, there was no nearby field in the
model which resembled the data here and so was left as is. Indeed,
comparing with the results of the Northern field it suggests that the data
here consists solely of Galactic components.

\subsubsection{Fields \bf$(350,+20)^\circ$}\label{350p20des}
\begin{figure}\centering
\includegraphics[width=82mm,angle=270]{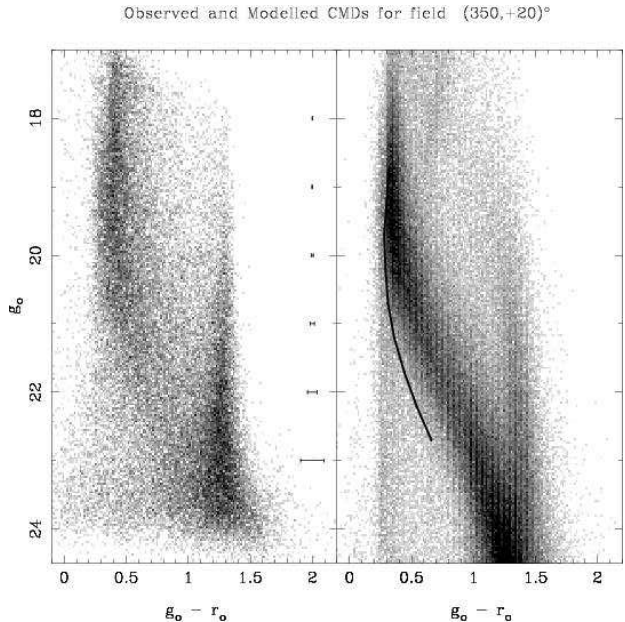}
\caption[]{Hess  plots of  ({\it l},{\it b}) = (350,+20)$\deg$ and its
  counterpart synthetic CMD. The synthetic CMD used here is actually
  (347.5,+20)$\deg$, the model field with the same coordinates is
  very similar to that seen in Figure~\ref{fig350m20}. It was found
  though that with a small shift in longitude, the model retrieves a
  CMD similar to the data. Since there is no reason to account for
  such a drastic change in the CMD in this direction, the
  (347.5,+20)$\deg$ field is used instead. The overlay here is at
  0.2 magnitudes of offset or 12.1 \kpc\ heliocentric distance. The
  overlay is fitted ``by-eye'' to the lower extremity of the dominant
  main sequence and is a good match to the model.
\label{fig350p20}}
\end{figure}

As for the Southern field at this Galactic longitude, this field also
had a problematic model CMD. However, it was noticed that a slight change
in coordinates in the model produced a CMD much more similar to the
data. So for this field, the comparison field is (347.5,+20)$\deg$
rather than (350,+20)$\deg$ (Figure~\ref{fig350p20}). The (347.5,+20)$\deg$ synthetic field is used
here due to its similarity with the data. The contrast between the
data and the model for the field in the South is deemed a glitch rather than a flaw in the entire model. The overlay here is
placed at 0.2 magnitudes and corresponds to the fainter edge of the
main sequence. It can be found at a heliocentric distance of 12.1
\kpc, although, as with the Southern field, the main sequence in
the model does seem to be stronger than the data. On the whole though, they are much more
similar here than in previous fields.

\subsubsection{Fields {\bf$(025,-20)^\circ$}}\label{025m20des}
\begin{figure}\centering
\includegraphics[width=82mm,angle=270]{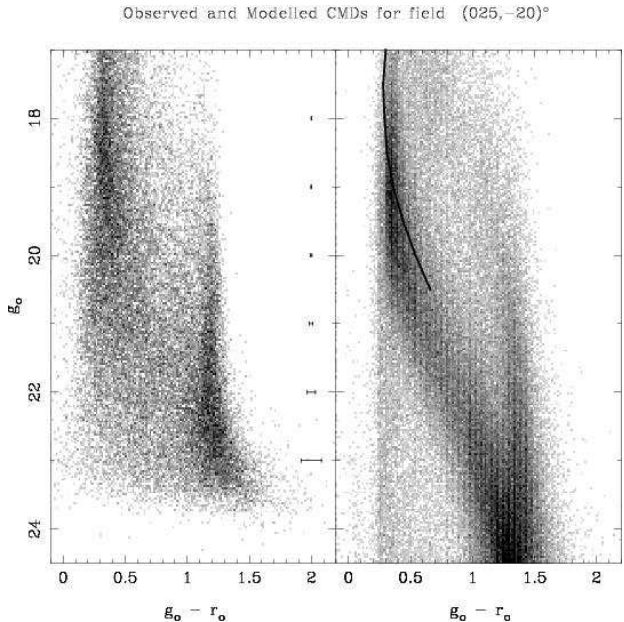}
\caption[]{Hess  plots of  ({\it l},{\it b}) = (025,-20)$\deg$ and its
  counterpart synthetic CMD. As for Figure~\ref{fig280m15}. The
  overlay is placed at an offset of -2.0, aligning with the bright end
  of the Milky Way main sequence feature in the data. The D$_{\odot}$
  is 4.4 \kpc.
\label{fig025m20}}
\end{figure}

This field (Figure~\ref{fig025m20}) completes the Monoceros Ring survey below the Galactic
plane which began with the INT/WFC survey. Despite having seeing of
typically 2$''$, the limiting magnitude of the data is still relatively
deep. In comparison with the model, the strong main sequence is
conspicuously missing from the data. In an attempt to compare the
features, the bright end of the weak Milky Way main sequence in the data
has been fit with an offset of -2.0 magnitudes. This converts to a
distance estimate of 4.4 \kpc\ which is reasonable match with the
model. It is uncertain why this field lacks a strong Thin Disc
presence in the data.

\subsubsection{Fields {\bf$(025,+20)^\circ$}}\label{025p20des}
\begin{figure}[!h]\centering
\includegraphics[width=82mm,angle=270]{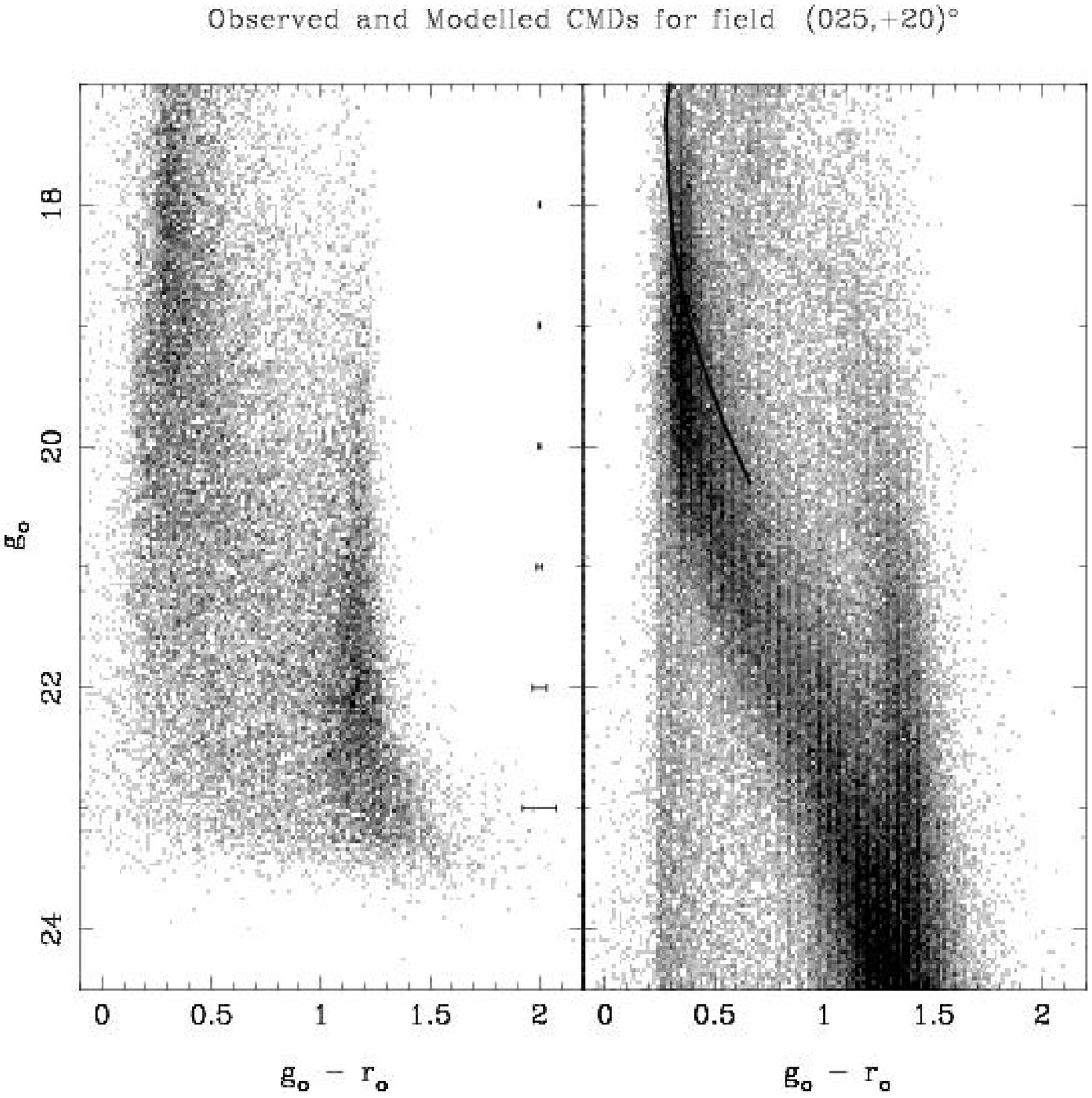}
\caption[]{ Hess  plots of  ({\it l},{\it b}) = (025,+20)$\deg$
  (Left) and its counterpart synthetic CMD (Right). As for
  Figure~\ref{fig280m15}. Despite the lack of a strong main sequence
  in the data, a main sequence overlay is fitted to what is estimated
  as the bright end of the Milky Way main sequence. This is at a
  magnitude offset of -2.2 or D$_{\odot} \sim$4.0 \kpc, in rough
  accordance with the model.
\label{fig025p20}}
\end{figure}

The final field  of the survey on the Northern side of the plane (Figure~\ref{fig025p20}) is
remarkably similar to its Southern counterpart. Again the model
predicts strong main sequence for the Thin Disc component which is not
present in the data. To provide some point of comparison the
approximate bright end of the weak Milky Way main sequence has been
estimated and is found at an offset of -2.2 magnitudes or 4.0
\kpc. The model predicts this edge here too. There is no evidence of the
Monoceros Ring in this field.

\section{Discussion}\label{discussion}

To date there are only two numerical simulations of the Monoceros Ring
and Canis Major structures, these are from \citet{2005MNRAS.362..906M}
and \citet{2005ApJ...626..128P}. The primary difference between these
two models is that the \citet{2005MNRAS.362..906M} model uses the properties
of the Canis Major overdensity as its constraints and the
\citet{2005ApJ...626..128P} model uses the data collected on the Monoceros Ring
up to that time. The following two sections compare the findings of this paper,
Paper I and the INT/WFC paper \citep{2005MNRAS.362..475C} with these
models. To make the comparison meaningful, in the next
sections the MRi is assumed to be a tidal stream. 

\subsection{Comparing the observations with the
  \citet{2005MNRAS.362..906M} model}
\begin{figure*}\centering 
\includegraphics[width=135mm,angle=270]{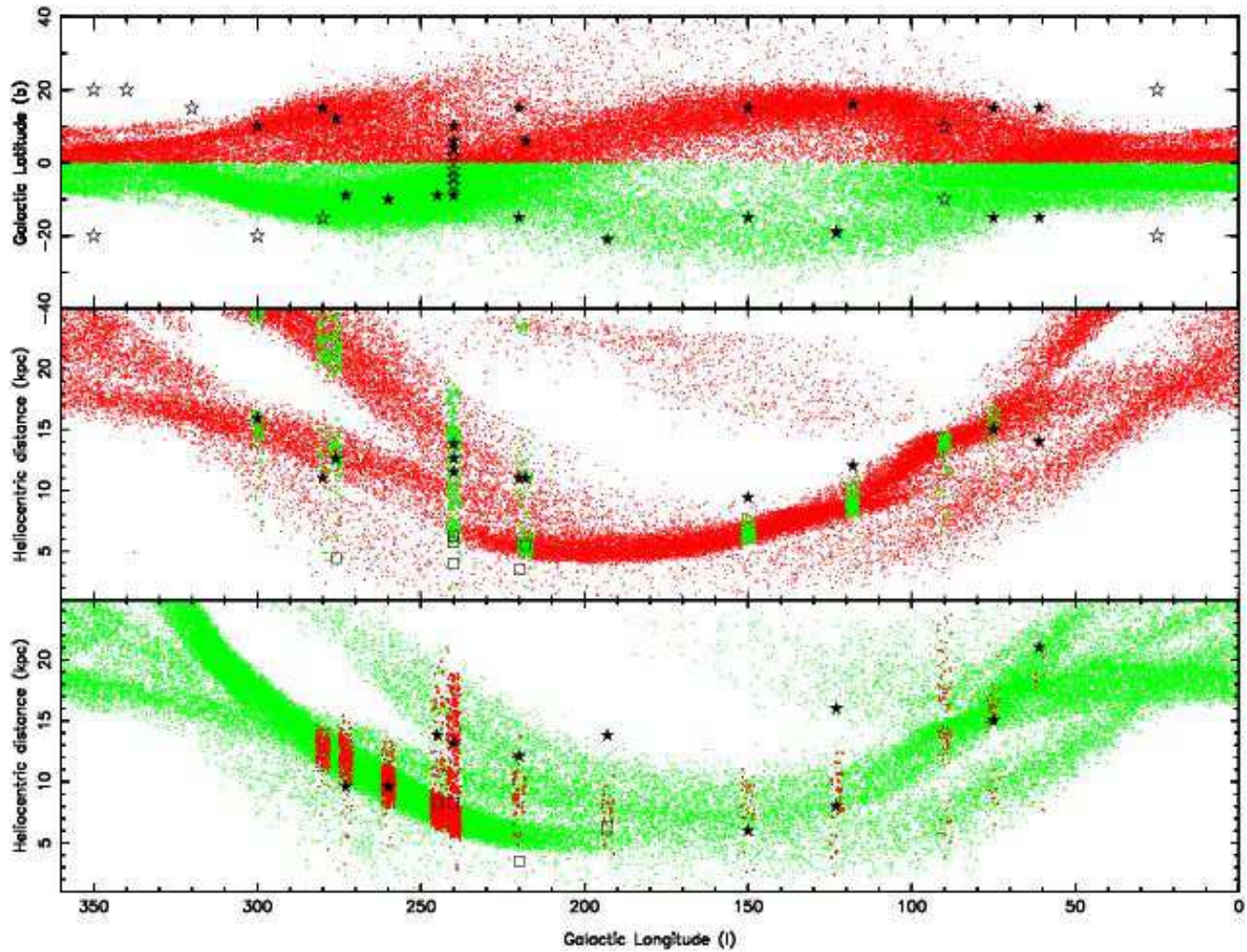}
\caption[]{Comparison of the \citep{2005MNRAS.362..906M} numerical simulation of the Monoceros
  Ring/Canis Major streams and the locations and distances of the
  detections (including tentative ones) arising from the survey. The top panel shows the
  simulation in Galactic coordinates, the centre panel shows only
  those fields and points from the model above the Galactic Plane
  against Heliocentric distance and the lower
  panel is for those points below the Galactic Plane. In the centre
  and lower panels, the points with the opposite colour (\ie green
  instead of red or vice versa) are the
  predicted distance of the model in the direction of the observed
  fields. The filled stars show fields with detections of the MRi and
  empty stars show fields in which the MRi was not detected. Empty
  squares show the location of the proposed CMa feature at that longitude as per the findings of Paper I.
\label{figmartincomp}}
\end{figure*}

The numerical simulation of \citet{2005MNRAS.362..906M} has been plotted
with the results from the entire survey (this paper, Paper I and
\citep{2005MNRAS.362..475C}) in Figure~\ref{figmartincomp}. The top panel shows the model in ({\it
  l,b}) space dividing the points, by colour, for those above and
below the Galactic Plane. All of the fields from the three papers have
been overplotted as either full or open stars. Full stars represent
fields with a Monoceros Ring detection and open stars are fields
without a Monoceros Ring detection. Tentative detections have been
included in this figure.

The middle panel contains only the points above the Plane plotted
against Heliocentric distance. For each field, the prediction of the
model for that location ({\it l,b}), is shown in green. This then
allows direct comparison between the finding of the survey with the
prediction of the model. To avoid clutter, the top panel only showed
MRi detections but for completeness the CMa detections from Paper I,
which reside in the same fields, are plotted as open squares. The
fields between {\it l} = (200 - 300)$\deg$ do seem to correspond well
to the model although there are a spread of distances which are
possible. The fields at ({\it l,b}) = (118,+16)$\deg$ and
(150,+15)$\deg$ are at distances greater than the predicted location
but they do vary in-step with the model and so could just represent
the model stream being too close heliocentrically. At ({\it l,b}) =
(90,+10)$\deg$ there is a conspicuous absence of the MRi. While in
other fields the overall data quality or area covered could be a
reason for a non-detection, but here there is no such problem. It is
unclear why the feature is absent. For the ({\it l,b}) =
(75,+15)$\deg$ field the detection again matches the model while the detection at ({\it
l,b}) = (61,+15)$\deg$ does not correspond well. The reasons for this is also uncertain.

The lower panel shows the predictions for the stream model below the
Plane. Around {\it l} = (240 - 276)$\deg$ the detections do roughly
correspond with the model and from {\it l} = (60 - 240)$\deg$ the
connection is more or less correlated with the general direction of
the stream. The two interesting omissions in
the South are ({\it l,b}) = (90,-10)$\deg$ and ({\it l,b}) =
(280,-15)$\deg$ as both these fields were expected to have MRi
components. As per the Northern field at {\it l} = 90$\deg$, the data
quality in its Southern counterpart field is sufficiently high to
robustly conclude no MRi feature is present here. For the {\it l} =
280$\deg$ field, the limiting magnitude is the second worst in the
sample but given that the predicted distance is more or less that of
the original detection by \citep{2002ApJ...569..245N} it should be
visible. The stream therefore does not pass through this field at the
distances suggested by the model.

Non-detections of the stream also provide an opportunity to test the
model. In all but a few cases the non-detections in the data are
supported as non-detection regions in the model. The survey is
too sparse to draw conclusions as to a potential path for the stream
but it does serve as a basis for future studies and models.

\subsection{Comparing the observations with the
  \citet{2005ApJ...626..128P} model}
\begin{figure*}\centering 
\includegraphics[width=135mm,angle=270]{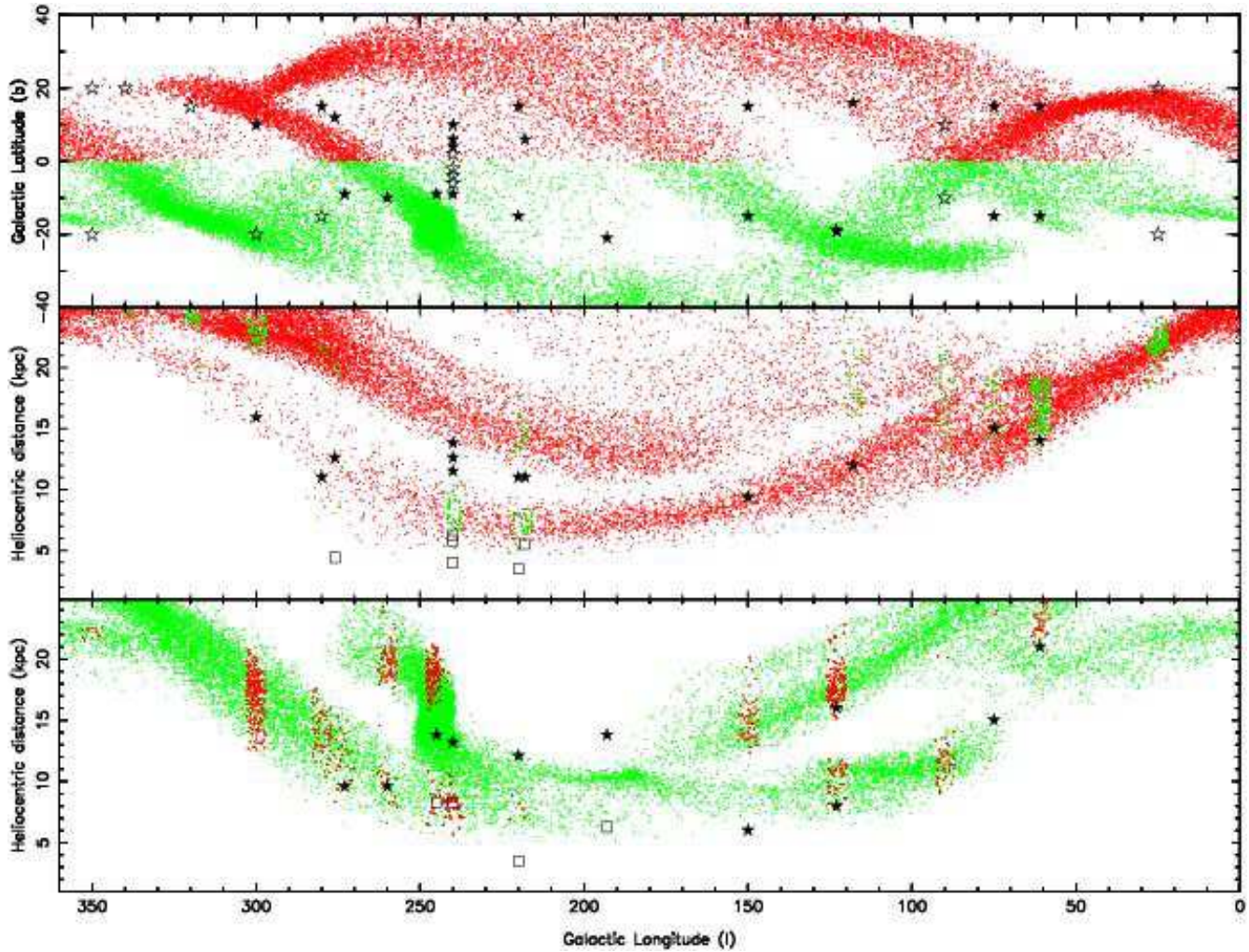}
\caption[]{As for Figure~\ref{figmartincomp}, but using the numerical
  simulation of \citet{2005ApJ...626..128P}. This model is useful for
  comparison as it uses the Monoceros Ring detections known at that
  time as
  constraints, rather than the overdensity of stars in the Canis Major
  region as used by \citet{2005MNRAS.362..906M}.
\label{figpenacomp}}
\end{figure*}

Interpreting the predictions of the \citet{2005ApJ...626..128P} model
has been done in the same way as for the \citet{2005MNRAS.362..906M}
model, primarily by comparing the locations and distances of the
observed structures with the distances and locations as predicted by
the model (Figure~\ref{figpenacomp}). The lower two panels show the
predicted stream locations from the model in each of the regions
surveyed.  Given that the \citet{2005ApJ...626..128P} model uses fewer particles
than the \citet{2005MNRAS.362..906M} model; a slightly bigger
area has been chosen around each field to sample enough model data
points. The correspondence with data is seemingly poorer for the
\citet{2005ApJ...626..128P} model and several non-detection regions
are supposedly populated by the stream.

For the Northern fields, many of the distances do seem to match the
predictions of the model. Close inspection of the model shows that the
detections are located on the wrong arm. Most of the fields observed
are located in sparsely populated regions of the model and do not
probe the predicted path of the model to higher
latitudes. The field at {\it l} = 25$\deg$ is possibly undetected due
to the predicted distance of the stream here. As discussed in $\S$~
\ref{analysis}, it is estimated that the technique used is only
sensitive to objects less than 20 kpc distant. The fields at ({\it
  l,b}) = (61 - 75)$\deg$ could be seen as confirmation of the stream
however the non-detection at ({\it l,b})) = (90,$+$10)$\deg$ is
difficult to explain. At ({\it l,b})) = (118,$+$16)$\deg$, the detection
is at least 5 kpc closer than the distance estimate from the
model. For the field centred on ({\it l,b})) = (150,$+$15)$\deg$, it
resides in an almost empty region of the model but seemingly the
detected stream here corresponds with the tidal arm at higher
latitudes. The discrepancy for this model around the {\it l} = 240$\deg$ region
is known and has been commented on by other authors. The fields at
longitudes {\it l} = (260 - 360)$\deg$ are simply unable to observe
the stream according to the model. The latitude for these fields is
not so much a problem and detections reported in this paper do not
match the model at all.

In the South, the match with the model is good around {\it l} =
60$\deg$, 123$\deg$ and close to the Plane around 250$\deg$. The
remaining fields occupy regions of low density in the model. A few
fields, like those at {\it l} = 90$\deg$ present real discrepancies
with the model. The main difference between the Northern
fields is that most of the predicted stream locations in the South
should have been detectable by the technique used here. In favour of
the model though, a significant proportion of the model is not sampled
in the survey as it is above a latitude of 20$\deg$. Finally, in
comparison to the original data used to create the model (see Figure
2, \citet{2005ApJ...626..128P}), most of the data points there reside
between {\it l} = (110 - 240)$\deg$. This corresponds to a relatively
sparse sampling in this survey. Another comparison of
this model against the available data is presented in
\citet{2007astro.ph..3601M}.

\subsection{Key locations to test the models}\label{locations}
Each model predicts that in key areas of the Galaxy there are
significant changes in the stream which could be used to both test the
model and provide further support to the tidal stream scenario as a
whole. In particular, these are the regions south of the purported
Canis Major dwarf galaxy over the longitude range {\it l} = (200 -
250)$\deg$ since in this location the two models predict different
approaches for the stream into the core. The region {\it l} = (130 -
220)$\deg$ is where the \citet{2005MNRAS.362..906M} model predicts the
leading tidal arm of the dwarf galaxy should decrease in latitude and
enter into the Disc. Finally the region {\it l} = (025 - 050)$\deg$:
here the stream is predicted to be close to the Plane ({\it b} =
$\pm$10$\deg$) in the \citet{2005MNRAS.362..906M} model and away from
the Plane ({\it b} = $\pm$20$\deg$) in the \citet{2005ApJ...626..128P}
model. Knowledge of the stream around the Bulge is needed to constrain
its position in all four quadrants of the Galaxy. The Bulge presents
an additional challenge in that the distance of the targets and the
high density of foreground stars will make the MRi difficult to
detect. The current dataset is unable to detect the MRi at the
distances predicted by the \citet{2005ApJ...626..128P}
model (25 - 30 kpc) and the \citet{2005MNRAS.362..906M} model
predictions of 15 - 25 kpc are yet to be tested so close the Plane.

\subsection{Insights into the nature of the Galactic Warp}
One of the key properties of the Galactic Disc is the Warp. Around
{\it l} = 90$\deg$ the Disc curves up from the {\it b} = 0$\deg$
position and around {\it l} = 270$\deg$ it curves down. Studies into
the putative Canis Major dwarf galaxy have had to contend with the
close proximity of the Warp and much debate has centred on whether the
CMDs in this region can be fully explained by the Warp or require an
additional source of stars. This part of the survey provides an
opportunity to understand the impact of the Galactic Warp on CMDs,
through a closer inspection of fields ({\it l,b}) =
(280,$\pm$15)$\deg$
(Figures~\ref{fig280m15},\ref{fig280p15}). At {\it l} = 280$\deg$
these fields are very near the maxima of the Southern Warp. Firstly,
both these fields are well matched by the model and the Warp is seen clearly as an
excess of stars in the Southern field. This manifest as both as a
general increase in star counts and an obvious thickening of the Thin
and Thick Disc components as seen in the Southern field. How to identify
the different components of the Galaxy in the CMDs is shown in
Figure~2 of Paper I. Secondly, we see that the influence of the
Galactic Warp does not change the shape of the CMD. The Northern field
at {\it l} = 280$\deg$ is essentially a shifted version of its
Southern counterpart. This is important because a comparison of the
almost symmetric fields ({\it l,b}) = (240,+10)$\deg$ and ({\it l,b})
= (240,-9)$\deg$ (See Figure~\ref{figmonster} or Figures~10 and 17 from Paper
I.) is remarkably different. While the fields at {\it l}
= 280$\deg$ have more of a sharp edge to Thin-Thick Disc boundary the Canis
Major field shows a true curving Main Sequence which is unmatched in
the Northern field. Since the {\it l} = 280$\deg$ fields show that the
Warp does not seem to have an impact on the shape of the CMD, the
fields in Canis Major must be considered anomalous to the usual
Galactic Warp scenario. Whether this anomaly is caused by a dwarf galaxy is
uncertain, however these qualitative differences in the CMDs should be
investigated so that our understanding of this region is more complete.

\section{Conclusion}\label{conclusion}
This paper reports on 2 new detections and 7 non-detections of the
Monoceros Ring tidal stream. The results presented here conclude a
survey tracing this feature around the entire Galactic plane. The
previously reported detections of the survey are presented in
Figure~\ref{figmonster}. Comparing the relative strengths of the MRi
and the main MW population it appears qualitatively that the stream is
denser and broader above the Plane than below but as such there is no explanation why this
would be the case. The part of the overall survey presented here shows
no evidence of the strong Canis Major dwarf main sequence in the
CMDs. The CMa sequence is historically identified as a shift in the
position and shape of the strongest main sequence in the CMD. For the
fields presented here, the dominant main sequence feature in the CMDs
is easily attributed to the Thin and Thick Discs. In each instance
where the distance has been determined to these structures it is in
accordance with the Besan\c{c}on synthetic galaxy model
predictions. Therefore they can be confidentially associated with the
MW. The only field which might be expected, from the
\citet{2005MNRAS.362..906M} model, to contain the CMa
signature is ({\it l,b}) = (280,-15)$\deg$. This field does not show
this CMa-style sequence in the CMD (Figure~\ref{fig280m15}).

Comparing these new MRi detections with the two current numerical
simulations of the stream and putative dwarf galaxy progenitor, has
led to inconclusive results. The \citet{2005MNRAS.362..906M} model
north of the Galactic Plane roughly traces the locations of the
detections. In the South the correspondence
between the model and the detections is adequate with some noted
exceptions. Several detections presented in this survey indicate, with
reasonable certainty, the locations in which the model is incorrect. For the
\citet{2005ApJ...626..128P} model, there is less correlation
between the data points and the predicted stream locations than is
seen with the \citet{2005MNRAS.362..906M} model. Although some points
do seem to represent a better fit it is important to note though that
a significant proportion of the \citet{2005ApJ...626..128P} model does
reside outside of a Galactic latitude of {\it b} = $\pm$20$\deg$. So
much of the model has not been sampled by this survey. Indeed, it is easy to see that this
survey is too narrow in Galactic latitude in comparison with the data
used to construct the model and the predictions it makes. Drawing a conclusion based on
these results in inadvisable but there is little here to strongly support this
model. Both models obviously will require reworking to include the new
information available along with more observations to test their predictions.

With regard to the Besan\c{c}on synthetic galaxy model, there is no
presence of the MRi as part of natural Galactic structure. In almost
all fields in this survey, the bulk Milky Way components of Thin,
Thick Disc have been accurately modelled. There is no systematic
discrepancy between the model and data even in regions containing the
Galactic Warp. Only the regions around Canis Major, as discussed in
Paper I, show a definite shift from the observational data. Given the
data supports the predictions of the Besan\c{c}on model in all but the
MRi detections, it is reasonable to assume this structure is indeed
additional to the usual Galactic components. 

Determining the density profile of this feature around the Galaxy and
indeed connecting detections is an important next step in resolving its origins. To
date, targeted deep surveys, such as this, have resolved many important questions
surrounding this structure. This survey sheds some light on
the impact of the Galactic Warp on the Colour-Magnitude Diagrams
showing it does not effect its morphology significantly and that the
Besan\c{c}on model is adequate for most fields. This has implications with
regard how the fields in the Canis Major region are to be interpreted
as the fields there have obviously different characteristics. While
the nature of the Monoceros Ring still remains quite elusive, this is primarily due
to its large extent on the sky and its location close to the
Plane. For the time being, both the Galactic origin scenario and the
tidal stream hypothesis are still possibilities for this
structure. The completed survey, presented here, has shown that a
targeted campaign of observations can provide insights on not only this
structure but also generic Galactic structures as well.

\section{Acknowledgments}
M.B. acknowledges the financial support of INAF to this research through the 
grants PRIN05 - CRA 1.06.08.02 and PRIN07 - CRA 1.06.10.04. RRL would like to thank LKN for her on-going support.
BCC thanks the referee for their constructive comments and B. Carry
for his work with the lighting.

\newcommand{\aap}{A\&A}
\newcommand{\apj}{ApJ}
\newcommand{\apjl}{ApJ}
\newcommand{\aaps}{AAPS}
\newcommand{\aj}{AJ}
\newcommand{\mnras}{MNRAS}
\newcommand{\nat}{Nature}

\bsp

\label{lastpage}

\end{document}